\documentclass[10pt,superscriptaddress,pra,twocolumn]{revtex4}%
\usepackage{amsfonts}
\usepackage{amsmath}
\usepackage{amssymb}
\usepackage{graphicx}
\usepackage{graphics}
\usepackage[usenames]{color}%

\begin{document}
\title{\textbf{Numerical study of three-body recombination \\ for systems with many bound states}}
\author{Jia Wang}
\affiliation{Department of Physics and JILA, University of Colorado,
Boulder, CO 80309}
\author{J.P. D'Incao}
\affiliation{Department of Physics and JILA, University of Colorado,
Boulder, CO 80309}
\author{Chris H.~Greene}
\affiliation{Department of Physics and JILA, University of Colorado,
Boulder, CO 80309}
\begin{abstract}
Three-body recombination processes are treated numerically for a system of three identical bosons. The two-body model potentials utilized support many bound states, a major leap in complexity that produces an intricate structure of sharp nonadiabatic avoided crossings in the three-body hyperradial adiabatic potentials at short distances. This model thus displays the usual difficulties of more realistic systems. To overcome the numerical challenges associated with sharp avoided crossings, the slow variable discretization (SVD) approach is adopted in the region of small hyperradii. At larger hyperradii, where the adiabatic potentials and couplings are smooth, the traditional adiabatic method suffices. Despite the high degree of complexity, recombination into deeply bound states behaves regularly due to the dominance of one decay pathway, resulting from the strong coupling between different recombination channels. Moreover, the usual Wigner threshold law must be modified for excited incident recombination channels.
\end{abstract}
\maketitle
\section{Introduction}
Three-body recombination has attracted much theoretical and experimental research interest in recent years. Recombination is the process in which three free particles collide to form a two-body state and a free particle, with the released kinetic energy being distributed between the final collisional partners. Such reactions are common and important in chemical reactions and in atomic, molecular, and nuclear physics. In ultracold degenerate Fermi gases \cite{K3ExpFermions} recombination has been used as a process to form weakly bound diatomic states, crucial for the experimental realization of the BEC-BCS crossover physics. In fact, it was shown in Ref.~\cite{SingleScat} that the use of recombination as an efficient way to produce weakly bound diatomic molecules can be extended to systems other than fermionic gases.  For colliding BECs at precisely-tuned relative velocity, the formation of molecules via 3-body recombination can also be used to form molecules efficiently owing to a double Bose enhancement \cite{BorcaPRL2003}. Recombination processes normally release a high amount of kinetic energy, producing atomic losses that often limit the lifetimes of Bose-Einstein condensates (BEC) \cite{K3ExpBosons}. Moreover, three-body recombination has been recognized as one of the most important scattering observables in which features related to the universal Efimov physics can be manifested \cite{Nielsen1999,Esry1999,BraatenReview,KnMehta}. Near a two-body Feshbach resonance, i.e., when the $s$-wave scattering length $a$ is much larger than the range $r_{0}$ of the interactions, Efimov states can occur, causing interference and resonant effects in recombination. The experimental observation of these features in recombination has been recently used as evidence of Efimov physics in ultracold quantum gases \cite{FerlainoPhysics}.

From the theoretical viewpoint, quantitative calculations of recombination for the typical alkali atoms used in experiments in ultracold gases are limited by the large number of diatomic states existing in such systems. Most of the available calculations for recombination for realistic systems have been confined to model systems possessing just a few bound states and/or systems with small scattering lengths, and even these are challenging calculations \cite{K3Calc}. As a result, recombination calculations relevant for ultracold gases can only be made in the universal regime $|a|\gg r_{0}$, by using simple potential models (with a few-bound states) or else by simply modeling the decay into all deeply bound molecular states through a single inelastic parameter \cite{BraatenReview}. However, in ultracold gases experiments the condition of universality is typically {\em not} deeply satisfied, making it desirable to perform more realistic calculations involving more sophisticated two-body models with, eventually, a larger number of deeply bound states. This paper develops a methodology within the hyperspherical adiabatic representation that permits the treatment of systems with many bound states.

The present study still utilizes two-body potentials models that are, however, designed to support many bound states, and therefore mimic three-body collisions for more realistic scenarios. In the hyperspherical representation, the existence of many bound states leads to a complex set of sharp nonadiabatic avoided crossings in the hyperspherical potential curves at short distances. The large number of sharp avoided crossings creates numerical difficulties for the traditional adiabatic representation as formulated with $d/dR$ couplings\cite{Suno2011}. To overcome these numerical difficulties, one solution is to use the slow variable discretiazation (SVD) method proposed by Tolstikhin \emph{et al} \cite{Tolstikhin1996}. The SVD method has been successfully applied to three-body bound-state calculations \cite{Suno2011,Jia2010} and three-body collisions for the $\rm{H+Ne_2}$ system \cite{Pack2002}. These calculations, however, did not require numerical solution of the hyperradial equation [see Eq.~(\ref{RadialEq}) below] out to large distances. To study ultracold collision processes such as recombination in the large scattering length limit, it is crucial to solve the hyperradial equation out to very large distances. Since application of SVD over the entire space is demanding in terms of memory and cpu-time, it is in fact much more efficient to separate the domain of hyperradii into two regimes. At short distances, where many avoided crossings appear, the SVD method is applied, while at large distances, where the adiabatic potential curves are smooth, the traditional adiabatic method \cite{BurkeThesis} is utilized.

This two-pronged strategy enables efficient calculation of the three-body recombination rate at low collision energy, with extremely stable results for a two-body potential model supporting up to about 10 bound states. This numerical capability of calculating recombination with many bound states permits us to study the final state distribution of the recombination rate, $K_{3}$. One unexpectedly simple finding is that the branching ratio of recombination into a particular final ($f$) channel, defined as
\begin{equation}\label{ratio}
r_{3}^{\left( {f \leftarrow i} \right)}  = \frac{{K_3^{\left( {f \leftarrow i} \right)} }}{{\sum\limits_f {K_3^{\left( {f \leftarrow i} \right)} } }}
\end{equation}
is the {\em same} for different initial ($i$) three-body collision channels. In the above equation, $K_{3}^{(f\leftarrow i)}$ is the partial recombination from the initial three-body channel $i$ to a particular final channel $f$. The threshold laws for the partial recombination rates have also been considered, i.e., recombination processes occurring from excited three-body continua. These partial rates are observed to deviate from the usual Wigner threshold law. Specifically, the energy dependence of the partial recombination rates display a much weaker suppression than the usual Wigner analysis \cite{Esry2001,Dincao2005} for excited continuum channels. These numerical results can be interpreted as the manifestation of a strong coupling between three-body continuum channels. This is further quantified through a perturbation series expansion of the scattering matrix that reveals the three-body recombination pathways  at low collision energies.

This article is organized as follows. Section II discusses the numerical methods adopted in this study. Section III shows the numerical results for the three-body recombination rates , and Section IV presents our analysis of the recombination pathways. Section V summarizes and concludes.

\section{Method}
The system studied here consists of three identical bosons with masses $m_1=m_2=m_3=m$. After separating out the center-of-mass motion, this triatomic system is described using modified Smith-Whitten hyperspherical coordinates \cite{Suno2002,WhittenSmith1968}: $\left\{{R,\Omega}\right\} \equiv \left\{{R,\theta,\varphi,\alpha,\beta,\gamma}\right\}$. $R$ is the hyperradius describing the overall size of the system, and the hyperangles $\theta$ and $\varphi$ describe the internal motion of the three-body system. $\alpha,\beta$ and $\gamma$ are the usual Euler angles. In these hyperspherical coordinates, the interparticle distances \cite{Suno2002} are defined as,
\begin{subequations}
\begin{eqnarray}
r_{12} &=& 3^{ - 1/4} R\left[ {1 + \sin \theta \sin \left( {\varphi  - \pi /6} \right)} \right]^{1/2},\\
r_{23} &=& 3^{ - 1/4} R\left[ {1 + \sin \theta \sin \left( {\varphi  - 5\pi /6} \right)} \right]^{1/2},\\
r_{31} &=& 3^{ - 1/4} R\left[ {1 + \sin \theta \sin \left( {\varphi  + \pi /2} \right)} \right]^{1/2},
\end{eqnarray}
\end{subequations}
where the $i$ and $j$ indices in $r_{ij}$ label particles $i$ and $j$. The present study uses a two-body potential model potential in form of
\begin{equation}\label{sechsqpotential}
v\left( {r_{ij} } \right) = D{\rm{sech}}^2 \left( {\frac{{r_{ij} }}{{r_0 }}} \right),
\end{equation}
where the coefficient $D$ is negative, and its magnitude is chosen to be large enough to support $8$ to $10$ two-body bound states ($4$ to $5$ s-wave bound states).

In hyperspherical coordinates, the three-body Schr\"odinger equation is given (in atomic units, a.u.) by
\begin{widetext}
\begin{equation}\label{SchrodingEq}
\left[ { - \frac{{1 }}{{2\mu_{3b} }}\frac{{\partial ^2 }}{{\partial R^2 }} + \frac{{\left( {\Lambda ^2  + {15/4} }\right)  }}{{2\mu_{3b} R^2 }} + V\left( {R,\theta ,\varphi } \right) } \right]\psi_{\nu '} \left( {R,\Omega } \right) =  E \psi_{\nu '} \left( {R,\Omega } \right),
\end{equation}
\end{widetext}
where $\psi_{\nu '} = R^{5/2} \Psi_{\nu '}$ is the rescaled total wave function. The index $\nu '$ distinguishes different linearly independent solutions that are degenerate in energy and includes other ``exact'' quantum numbers. In Eq. (\ref{SchrodingEq}), $\Lambda ^2$ is the ``grand angular momentum operator'' \cite{ Suno2002}, $\mu_{3b}=\sqrt{m_1 m_2 m_3 / \left({m_1+m_2+m_3}\right)}=m/\sqrt{3}$ is the three-body reduced mass, and the total potential energy is defined as the pairwise sum of two-body interactions
\begin{equation}
V\left( {R,\theta ,\varphi } \right) = v\left( {r_{12} } \right) + v\left( {r_{23} } \right) + v\left( {r_{31} } \right).
\end{equation}

Eq. (\ref{SchrodingEq}) is solved in the hyperspherical adiabatic representation. Like the usual adiabatic approximation, the hyperspherical adiabatic potentials and channel functions are defined as solutions of the following adiabatic eigenvalue problem:
\begin{equation}\label{AdiabaticEq}
H_{\rm{ad}}\left({R, \Omega}\right) \Phi _\nu  \left( {R;\Omega } \right) = U_\nu  \left( R \right)\Phi _\nu  \left( {R;\Omega } \right),
\end{equation}
where the adiabatic Hamiltonian, containing {\em all} angular dependence and interactions, is defined as
\begin{equation}
H_{\rm{ad}}\left({R, \Omega}\right)=\left[ {\frac{{\Lambda ^2 }}{{2\mu_{3b} R^2 }} + \frac{{15}}{{8\mu_{3b} R^2 }} + V\left( {R,\theta ,\varphi } \right)} \right].
\end{equation}
Therefore, the adiabatic potentials and nonadiabatic couplings, obtained by solving Eq.~(\ref{AdiabaticEq}) for fixed values of $R$, contain all the correlations relevant to this problem.
Figures 1 and 2 show the typical adiabatic potential curves for the parameter $m = 7.2963 \times 10^{3}$, $D = -5.500\times 10^{-5}$ a.u. and $r_0 = 15$ a.u. Figure 1 exhibits several sharp nonadiabatic avoided crossings at short distances. Although in our representation sharp crossings are associated with vanishingly small transition probabilities, these avoided crossings can cause several numerical difficulties when solving for the hyperradial motion in the traditional adiabatic approach. As mentioned above, such difficulties are overcome by implementing the SVD method described in Ref.~\cite{Tolstikhin1996}. Figure 2 shows, however, that the adiabatic potentials at long distances are smooth and, therefore, are more suitable for traditional approaches.

The goal of our scattering study is to determine, from the solutions of Eq. (\ref{SchrodingEq}), the scattering matrix $\underline S$. In order to accomplish such goal, the R-matrix is first computed; this $\underline {\mathcal{R}}$ is a more fundamental quantity  that can be subsequently used to determine the scattering matrix [see Eqs. (\ref{Kmat}) and (\ref{Smat}) below]. As usual, the $R$-matrix $\underline {\mathcal{R}} \left( {R } \right)$ is defined as
\begin{equation}\label{RmatrixDef}
\underline {\mathcal{R}}  \left( {R } \right) = \underline F \left( R \right)\left[ {\underline {\widetilde F} \left( R \right)} \right]^{ - 1},
\end{equation}
where matrices $\underline F$ and $\underline {\widetilde F}$ are given in terms of the solutions of Eqs.~(\ref{SchrodingEq}) and (\ref{AdiabaticEq}) by:
\begin{subequations}\label{Fmatrix}
\begin{eqnarray}
F_{\nu \nu '} \left( R \right) &=& \int {d\Omega } \Phi _\nu  \left( {\Omega ;R} \right)^* \psi _{\nu '} \left( {\Omega ,R} \right), \\
\widetilde F_{\nu \nu '} \left( R \right) &=& \int {d\Omega \Phi _\nu  \left( {\Omega ;{R}} \right)} ^* \frac{\partial}{{\partial R}}\psi _{\nu '}  \left( {\Omega ,R} \right).
\end{eqnarray}
\end{subequations}

In the traditional adiabatic method, $\underline {\mathcal{R}} \left( {R } \right)$ is calculated by assuming an adiabatic expansion for the total wave function. In this case the $\nu'-th$ independent wave function at the specified energy $E$ is expanded in terms of the channel functions $\Phi_{\nu}(R;\Omega)$ with coefficients given by $F_{\nu\nu'}(R)$. After projection onto the channel functions $\Phi_{\nu}(R;\Omega)$ from Eq.~(\ref{SchrodingEq}), the resulting coupled system of ordinary differential equations can be solved for this $\nu'-$th independent solution:
\begin{eqnarray}\label{RadialEq}
&& \left[ { - \frac{1}{{2\mu_{3b}}}\frac{d}{{dR^2 }} + U_{\nu} \left( R \right)} -E \right]F_{\nu \nu '} \left( R \right)  \\ && - \frac{1}{{2\mu_{3b}}}\sum\limits_\mu  {\left[ {2P_{\nu \mu } \left( R \right)\frac{d}{{dR}} + Q_{\nu \mu } \left( R \right)} \right]} F_{\mu \nu '}\left( R \right)  = 0, \nonumber
\end{eqnarray}
with appropriated scattering boundary conditions for the hyperadial solutions $\underline{F}(R)$ and determine $\underline{\tilde{F}}(R)$ in terms of the derivatives of $\underline{F}(R)$ and $\Phi_{\nu}(R;\Omega)$, see Appendix A. The nonadiabatic couplings in Eq.~(\ref{RadialEq}) are defined as,
\begin{eqnarray}
P_{\nu \mu } \left( R \right) &=& \int {d \Omega } \Phi _\nu  \left( {R;\Omega } \right)^* \frac{\partial}{{\partial R}}\Phi _\mu  \left( {R;\Omega } \right), \\
Q_{\nu \mu } \left( R \right) &=& \int {d\Omega } \Phi _\nu  \left( {R;\Omega } \right)^* \frac{{\partial^2 }}{{\partial R^2 }}\Phi _\mu  \left( {R;\Omega } \right),
\end{eqnarray}
are the terms that control the inelastic transitions as well as the width of the resonances supported by the adiabatic potentials $U_{\nu}(R)$. Therefore, the accuracy of the results for three-body observables depends crucially on the accuracy of the calculated nonadiabatic couplings. Usually, these matrix elements are numerically evaluated by a simple differencing scheme, i.e.,
\begin{equation}
\frac{\partial }{{\partial R}}\Phi _\mu  \left( {R;\Omega } \right) \approx \frac{{\Phi _\mu  \left( {R + \Delta R;\Omega } \right) - \Phi _\mu  \left( {R- \Delta R;\Omega } \right)}}{{2\Delta R}}.
\end{equation}
This scheme works well, however,  only when $P_{\nu \mu } \left( R \right)$ and $Q_{\nu \mu } \left( R \right)$ are smooth function of $R$, e.g., at large distances and/or for systems with just a few bound states. Clearly, this scheme suffers from tremendous numerical difficulties arising from sharp nonadiabatic avoided crossings at short distances in systems with many bound states. In that case, the SVD approach offers a much more stable and accurate approach for solving Eq. (\ref{SchrodingEq}).

One key ingredient for implementing the SVD approach is the use of the discrete variable representation (DVR) \cite{Parker1982, McCurdy2000}. Our DVR basis functions $\pi_i\left({R}\right)$ are defined by the Gauss-Lobatto quadrature points $x_i$ and weights $w_i$ \cite{Wyatt1998}. This quadrature approximates integrals of a function $g\left( x \right)$ as
\begin{equation}
\int_{ - 1}^1 {g\left( x \right)} dx \cong \sum\limits_{i = 1}^N {g\left( {x_i } \right)} w_i.
\end{equation}
After scaling the quadrature points and weights, the above equation is generalized to treat definite integrals over an arbitrary interval $R \in \left[ {a_1 ,a_2 } \right]$:
\begin{equation}\label{LobattoQuadrature}
\int_{a_1 }^{a_2 } {g\left( R \right)dR}  \cong \sum\limits_{i = 1}^N {g\left( {R_i } \right)} \widetilde {w}_i,
\end{equation}
where
\begin{equation}
\widetilde {w}_i  = \frac{{a_2  - a_1 }}{2}w_i , R _i = \frac{{a_2  + a_1 }}{2}x_i  + \frac{{a_2  - a_1 }}{2}.
\end{equation}
Eq. (\ref{LobattoQuadrature}) is exact for polynomials whose degree is less than or equal to $2N-1$.  We construct the DVR basis functions as
\begin{equation}\label{DVR}
\pi _i \left( R \right) = \sqrt {\frac{1}{{\widetilde w_i }}} \prod\limits_{j \ne i}^N {\frac{{R - R_j }}{{R_i  - R_j }}},
\end{equation}
which have the important property that
\begin{equation}\label{DVRProperty}
\pi _i \left( {R_j } \right) = \sqrt {\frac{1}{{\widetilde w_i }}} \delta _{ij}.
\end{equation}
Hence, matrix elements of any function $H\left({R}\right)$ obey,
\begin{equation}
\int_{a_1 }^{a_2 } {\pi _i \left( R \right)H\left( R \right)\pi _j \left( R \right)dR}  \cong H\left( {R_i } \right)\delta _{ij},
\end{equation}
which is usually called the DVR approximation.

Next, the R-matrix propagation method is combined with the SVD approach, following the logic of Ref. \cite{Burke1982} and using the DVR basis given by Eq.~(\ref{DVR}). For a given R-matrix [Eq.~(\ref{RmatrixDef})] at $R = a_1$ one uses the R-matrix propagation method to calculate the corresponding R-matrix at another point $R = a_2$, as follows. The solution $\psi_{\nu '}$ is expanded in the radial DVR basis $\pi_j\left({R}\right)$ and in hyperangles in terms of the adiabatic hyperspherical channel functions as
\begin{equation}\label{SVDExpand}
\psi _{\nu '} \left( {R,\Omega } \right) = \sum\limits_{j\mu } {c_{j\mu,\nu' } \pi _j \left( R \right)\Phi _\mu  \left( {\Omega ;R_j } \right)},
\end{equation}
where $\Phi _\nu  \left( {\Omega ;R_j } \right)$ is the $\nu-$th hyperspherical adiabatic channel function calculated at $R=R_j$.

Substituting Eq. (\ref{SVDExpand}) into Eq. (\ref{Fmatrix}) yields the values of matrix elements of $F_{\nu \nu '}$ and $\widetilde F_{\nu \nu '}$ at $R=a_1$ and $R=a_2$ boundaries in terms of the coefficients of Eq. (\ref{SVDExpand}):
\begin{subequations}\label{FmatrixelementSVD}
\begin{eqnarray}
F_{\nu \nu '} \left( {a_1 } \right) &=& \sum\limits_{j } {c_{j\nu,\nu' } \pi _j \left( {a_1 } \right)} ,\\
F_{\nu \nu '} \left( {a_2 } \right) &=& \sum\limits_{j } {c_{j\nu,\nu' } \pi _j \left( {a_2 } \right)},\\
\widetilde {F}_{\nu \nu '} \left( {a_1 } \right) &=& \sum\limits_{j \mu} {c_{j\mu,\nu' } O_{\nu \mu }^{1j} \pi _j' \left( {a_1 } \right)} ,\\
\widetilde {F}_{\nu \nu '} \left( {a_2 } \right) &=& \sum\limits_{j \mu} {c_{j\mu,\nu' } O_{\nu \mu }^{Nj} \pi _j' \left( {a_2 } \right)} .
\end{eqnarray}
\end{subequations}
where $O_{\nu \mu}^ {i j}$ are the overlap matrix elements,
\begin{equation}
O_{\nu \mu }^{ji}  = \int {d\Omega \Phi _\nu  \left( {\Omega ;R_j } \right)} ^* \Phi _\mu  \left( {\Omega ;R_i } \right).
\end{equation}
Note that the determination of $\underline{F}$ and $\underline{\tilde{F}}$ according to the above expressions only depend on derivatives of the well-behaved DVR basis [$\pi_{j}'(R)$] . Therefore, this approach is much better suited to handle the complex structure of avoided crossings present in systems with multiple bound states (see Fig. 1).

Over an interval $R \in [a_1,a_2]$, the DVR approximation based on quadrature gives:
\begin{equation}
\int_{a_1}^{a_2}  {\pi _i \left( R \right)H_{\rm{ad}} \left( {R,\Omega } \right)\pi _j \left( R \right)dR \approx H_{\rm{ad}} \left( {R_i,\Omega } \right)\delta _{ij} }.
\end{equation}
Expansion of the Schr\"odinger equation in the same numerical basis functions as in Eq. (\ref{SVDExpand}) and integration by parts yields the equation for the expansion coefficients $c_{j\mu,\nu'}$ (in vector notation, $\vec c_{\nu'}$):
\begin{equation}\label{matrixform}
\left[{\widetilde H-E}\right]\vec c_{\nu'} = L\vec c_{\nu'},
\end{equation}
or, equivalently,
\begin{equation}\label{matrixform2}
\vec c_{\nu'} = \left[{\widetilde H-E}\right]^{-1}L\vec c_{\nu'}.
\end{equation}
Here the matrix elements of $\widetilde H$ and $L$ are given by,
\begin{eqnarray}
\widetilde H_{i\nu,j\mu} &=&   \frac{1}{{2\mu_{3b}}}\left[ {\int_{a_1 }^{a_2 } {\frac{{d\pi _i \left( R \right)}}{{dR}}\frac{{d\pi _j \left( R \right)}}{{dR}}dR} } \right]O_{\nu \mu }^{ij} \nonumber \\ && + U_\nu  \left( {R_i } \right) \delta _{\nu \mu } \delta _{ij}  , \\
\label{Lmatrixelements} L_{i\nu,j\mu} &=& {\frac{1}{{2\mu_{3b}}}} \left. {\left[ {\pi _i \left( R \right)\frac{{d\pi _j \left( R \right)}}{{dR}}O_{\nu \mu }^{ij} } \right]} \right|_{a_1 }^{a_2 } .
\end{eqnarray}
Diagonalizing $\widetilde H$ over the range $\left[{a_1 , a_2}\right]$ gives,
\begin{equation}\label{diagH}
\vec {x_n}^{T}  \widetilde H \vec{x_n'}  = \varepsilon_n \delta_{n,n'},
\end{equation}
and the completeness relation of $\vec x_n$,
\begin{equation}
\sum\limits_n {\vec x_n \vec x_n^T }  = \underline 1,
\end{equation}
where $\underline 1$ is an identity matrix. Eq. (\ref{matrixform2}) is then rewritten as
\begin{equation}\label{cvec}
\vec c_{\nu '}  = \left[ {\widetilde H - E} \right]^{ - 1} \sum\limits_n {\vec x_n \vec x_n^T } L\vec c_{\nu '}  = \sum\limits_n {\frac{{\vec x_n \vec x_n^T }}{{\varepsilon _n  - E}}} L\vec c_{\nu '}.
\end{equation}
Substitution of the matrix elements of $L$ from Eq. (\ref{Lmatrixelements}) and insertion of the definition of $F_{\nu \nu'}$ and $\widetilde F_{\nu \nu'}$ at $a_1$ and $a_2$ in Eq. (\ref{FmatrixelementSVD}), finally gives
\begin{subequations}
\begin{eqnarray}
F_{\nu \nu '} \left( {a_1 } \right) &=& \sum\limits_{n\mu } {\frac{{u_\nu ^{\left( n \right)} \left( {a_1 } \right)u_\mu ^{\left( n \right)} \left( {a_2 } \right)}}{{2\mu_{3b}\left( {\varepsilon_n  - E} \right)}}\widetilde F_{\mu \nu '} \left( {a_2 } \right)}  \nonumber \\ &&- \sum\limits_{n\mu } {\frac{{u_\nu ^{\left( n \right)} \left( {a_1 } \right)u_\mu ^{\left( n \right)} \left( {a_1 } \right)}}{{2\mu_{3b}\left( {\varepsilon_n  - E} \right)}}\widetilde F_{\mu \nu '} \left( {a_1 } \right)}, \\
F_{\nu \nu '} \left( {a_2 } \right) &=& \sum\limits_{n\mu } {\frac{{u_\nu ^{\left( n \right)} \left( {a_2 } \right)u_\mu ^{\left( n \right)} \left( {a_2 } \right)}}{{2\mu_{3b}\left( {\varepsilon_n  - E} \right)}}\widetilde F_{\mu \nu '} \left( {a_2 } \right)} \nonumber \\ &&- \sum\limits_{n\mu } {\frac{{u_\nu ^{\left( n \right)} \left( {a_2 } \right)u_\mu ^{\left( n \right)} \left( {a_1 } \right)}}{{2\mu_{3b}\left( {\varepsilon_n  - E} \right)}}\widetilde F_{\mu \nu '} \left( {a_1 } \right)},
\end{eqnarray}
\end{subequations}
where,
\begin{equation}
u_\nu ^{\left( n \right)} \left( R \right) = \sum\limits_j {x_{j\nu, n} \pi _j \left( R \right)},
\end{equation}
and $x_{j\nu, n}$ are elements of the vector $\vec x_n$.

Our next step introduces the following matrices
\begin{subequations}
\begin{eqnarray}
\left( {\mathcal{\underline  R}_{11} } \right)_{\nu \mu } &=& \sum\limits_{n\mu } {\frac{{u_\nu ^{\left( n \right)} \left( {a_1 } \right)u_\mu ^{\left( n \right)} \left( {a_1 } \right)}}{{2\mu_{3b}\left( {\varepsilon_n  - E} \right)}}} , \\
\left( {\mathcal{\underline R}_{12} } \right)_{\nu \mu }  &=& \sum\limits_{n\mu } {\frac{{u_\nu ^{\left( n \right)} \left( {a_1 } \right)u_\mu ^{\left( n \right)} \left( {a_2 } \right)}}{{2\mu_{3b}\left( {\varepsilon_n  - E} \right)}}}, \\
\left( {\mathcal{\underline R}_{21} } \right)_{\nu \mu }  &=& \sum\limits_{n\mu } {\frac{{u_\nu ^{\left( n \right)} \left( {a_2 } \right)u_\mu ^{\left( n \right)} \left( {a_1 } \right)}}{{2\mu_{3b}\left( {\varepsilon_n  - E} \right)}}}, \\
\left( {\mathcal{\underline R}_{22} } \right)_{\nu \mu }  &=& \sum\limits_{n\mu } {\frac{{u_\nu ^{\left( n \right)} \left( {a_2 } \right)u_\mu ^{\left( n \right)} \left( {a_2 } \right)}}{{2\mu_{3b}\left( {\varepsilon_n  - E} \right)}}}.
\end{eqnarray}
\end{subequations}
and after some manipulation, the matrix equation is finally obtained that determines the R-matrix propagation from $a_1$ to $a_2$:
\begin{equation}\label{RmatProb}
\mathcal{\underline R}\left( {a_2 } \right) = \mathcal{\underline R}_{22}  - \mathcal{\underline R}_{21} \left[ {\mathcal{\underline R}_{11}  + \mathcal{\underline R}\left( {a_1 } \right)} \right]^{ - 1} \mathcal{\underline R}_{12}.
\end{equation}

In the SVD method, the overlap matrix $O_{\nu \mu }^{ji}$ requires us to calculate the channel functions $\Phi _\nu  \left( {\Omega ;R_j } \right)$ at every grid point $R_j$, which can be very memory demanding if one needs to perform calculations in a broad range of $R$. At large distances, therefore, we apply the traditional adiabatic approach combined with the R-matrix propagation method. In the traditional adiabatic method, the $P$ and $Q$ matrixes can be calculated on a sparse grid, and then interpolated and/or extrapolated on a much denser grid and larger distances. This strategy makes the calculation faster and it also requires less memory. The main difference between the traditional adiabatic approach and the SVD method is the use of a different three-body numerical basis. The details of this traditional approach and its connection with SVD method are discussed in Appendix A.

Once we have the R-matrix at large distances, the physical scattering matrix $\underline {\mathcal{S}}$ (and its close relative, the reaction matrix $\underline {\mathcal{K}}$) can be simply determined by applying asymptotic boundary conditions:
\begin{equation}\label{Kmat}
\underline {\mathcal{K}} = \left( {\underline f - \underline f'\underline {\mathcal{R}}} \right)\left( {\underline g - \underline g'\underline {\mathcal{R}}} \right)^{ - 1},
\end{equation}
\begin{equation}\label{Smat}
\underline {\mathcal{S}} = \left( {\underline 1 + i\underline {\mathcal{K}}} \right)\left( {\underline 1 - i\underline {\mathcal{K}}} \right)^{ - 1},
\end{equation}
where $\underline f$, $\underline f'$, $\underline g$ and $\underline g'$ are diagonal matrixes whose matrix elements are the energy-normalized asymptotic solutions $f_{\nu}$, $g_{\nu}$ and their derivatives $f_{\nu}'$, $g_{\nu}'$ respectively. $f_{\nu}$ and $g_{\nu}$ are given in terms of spherical Bessel functions: $f_\nu  \left( R \right) = \left( {2\mu_{3b}k_\nu  /\pi } \right)^{1/2} Rj_{l_\nu  } \left( {k_\nu  R} \right)$, $g_\nu  \left( R \right) = \left( {2\mu_{3b}k_\nu  /\pi } \right)^{1/2} Rn_{l_\nu  } \left( {k_\nu  R} \right)$, where $k_\nu$ and $l_\nu$ are determined by the asymptotic behavior of the potential in Eqs.~(\ref{Ufasym}-\ref{Uiasym}) \cite{addnote}. The three-body recombination rate is therefore given by \cite{Suno2002}
\begin{equation}
K_3^{\left( {f \leftarrow i} \right)}  = \frac{{192\pi^2\left( {2J + 1} \right)}}{{\mu_{3b}k^4 }}\left| {\mathcal{S}_{fi} } \right|^2,
\end{equation}
where $\mathcal{S}_{fi}$ is the appropriate S-matrix elements, $J$ is the total angular momentum of the system, and $k=\sqrt{2\mu_{3b}E}$ gives the hyperradial wave numbers in the incident channels.
\section{Three-body recombination rates}
\begin{figure}[t]
\includegraphics[width=0.5 \textwidth]{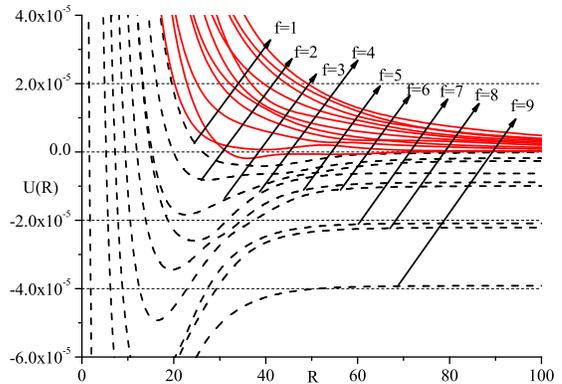}
\caption{(Color online) Adiabatic potential curves $U\left(R\right)$ at short distances $R$. Red solid lines represent the three-body continuum channels, i.e., the initial channels for recombination processes, and black dashed lines represent the final recombination channels.}
\end{figure}

\begin{figure}
\includegraphics[width=0.5 \textwidth]{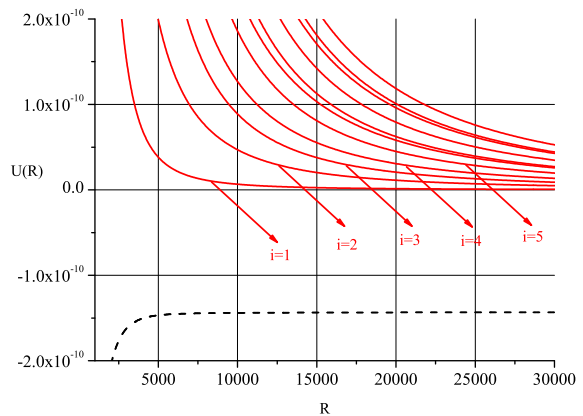}
\caption{(Color online) Same as Fig. 1, but for the adiabatic potential curves $U\left(R\right)$ at large distances $R$. This figure contrasts with Fig. 1 in the characteristically smooth behavior at large distances.}
\end{figure}

\begin{figure}[t]
\includegraphics[width=0.5 \textwidth]{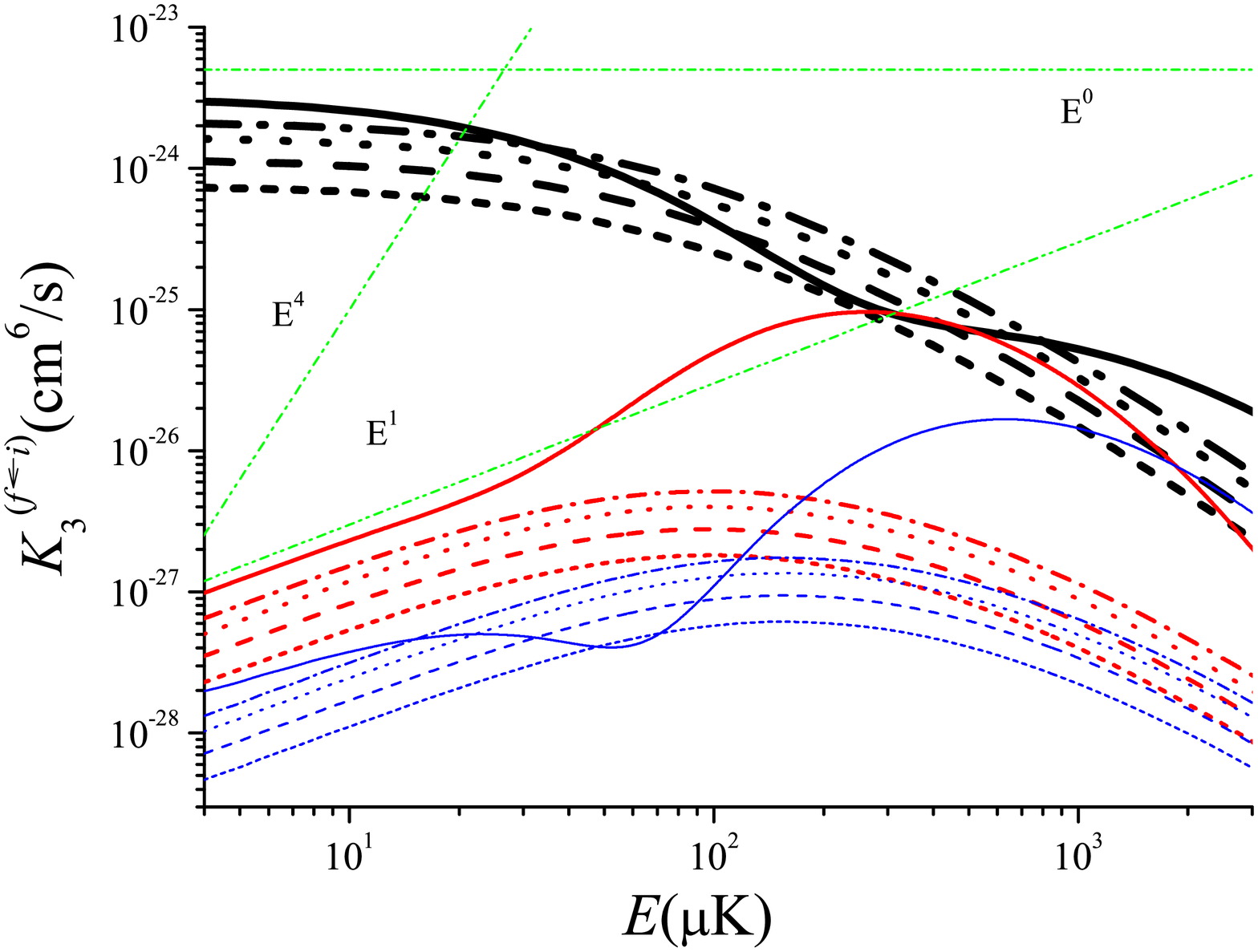}
\caption{(Color online) Partial recombination rate $K_3^{\left( {f \leftarrow i} \right)}$ as a function of the collision energy $E$ for the positive-scattering length case. The solid, dashed, dot, dash-dot and short dashed lines indicate partial recombination rates to different recombination channels: $f=1,2,3,4,5$, respectively. The thick black lines, thinner red lines and thinnest blue lines indicate recombination rates from different incident channels: $i=1,2,3$. The three green dash-dot-dot lines are proportional to $E^0$, $E^1$ and $E^4$ as indicated in the figure.}
\end{figure}

\begin{figure}
\includegraphics[width=0.5 \textwidth]{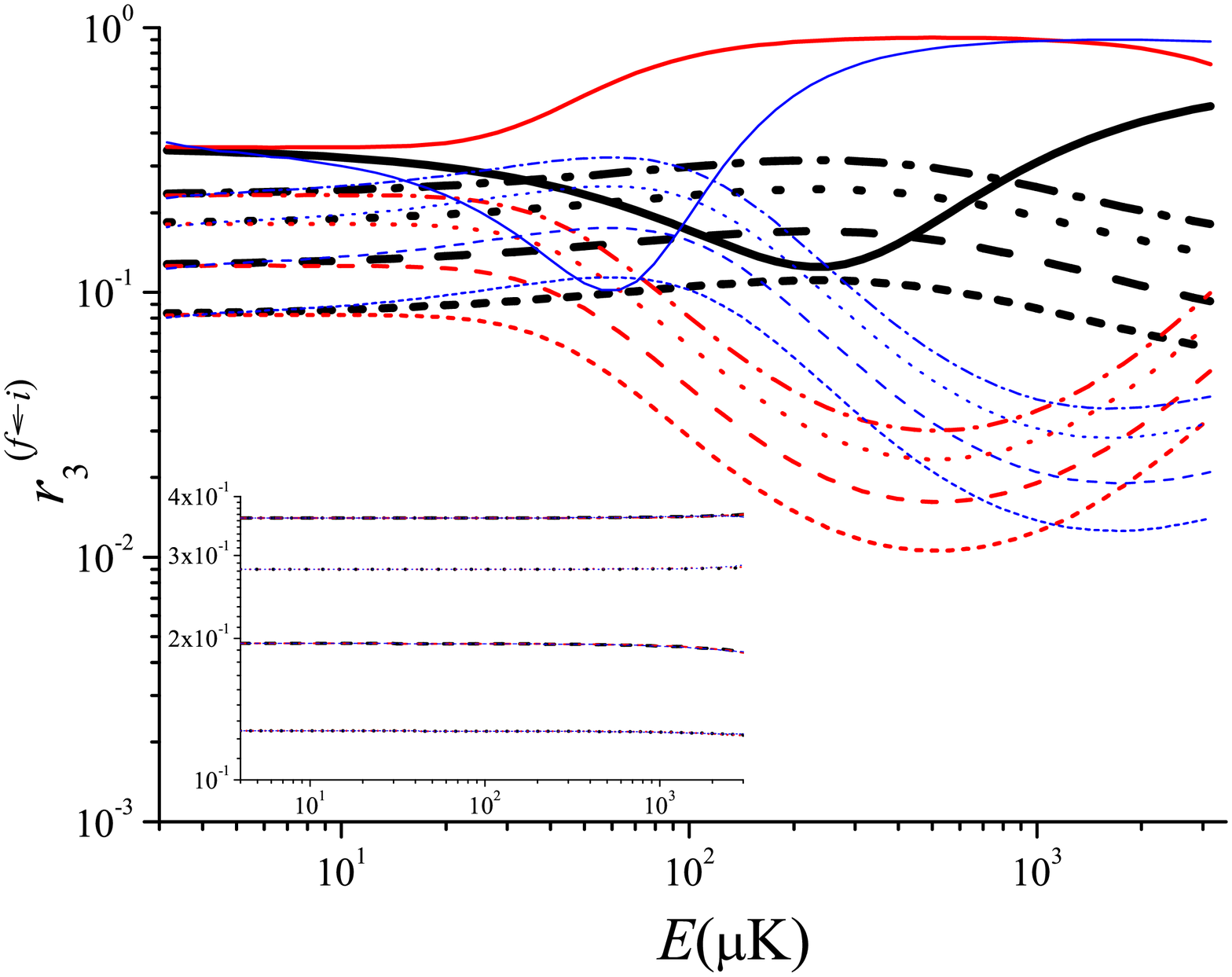}
\caption{(Color online) Branching ratio of the calculated recombination rates $r_3^{\left( {f \leftarrow i} \right)}$ as functions of the collision energy $E$ for the positive-scattering length case. The line-styles solid, dashed, dot, dash-dot and short-dashed indicate recombination rate to different recombination channels: $f=1,2,3,4,5$, respectively. The thick black lines, thinner red lines and thinnest blue lines indicate recombination rates from three different incident channels: $i=1,2,3$. The inset shows the branching ratio between the deep bound states only and it excludes the shallowest bound state (see the text for further details).}
\end{figure}

The present numerical study focuses on systems of three identical bosons with total angular momentum $J=0$, with parameters adjusted to represent the $^4\rm{He}$ system ($m_1=m_2=m_3=7.2963\times 10^{3}$ a.u. and $r_{0}=15$ a.u.). The two-body potential depth $D$ [see Eq.~(\ref{sechsqpotential})] is adjusted to tune the scattering length $a$ and explore both the positive- and negative-scattering length cases while supporting 8-10 bound states. The recombination rate near the unitary regime ($k\left| a \right| \approx 1$), is explored next for two sets of typical parameters: $D=-5.500 \times 10^{-5}$ a.u. for the positive-scattering length case, $a=1020.36$ a.u.; $D=-5.467 \times 10^{-5}$ a.u. for negative-scattering length case, $a=-1096.07$ a.u. For both cases $a$ is much larger than $r_{0}$ ($|a|/r_{0}\approx 70$) and therefore such calculations are solidly within the universal regime \cite{BraatenReview,Dincao2005}.

The black dashed lines in Figs. 1 and 2 denote the recombination channels, i.e., the final state channels of the recombination process. The effective hyperradial potentials $\widetilde U_\nu \left( R \right) \equiv U_\nu  \left( R \right) - Q_{\nu \nu } /\left( {2\mu _{3b} } \right)$ for these channels have asymptotic behavior given by
\begin{equation}\label{Ufasym}
\widetilde U_f  \left( R \right) \stackrel{R\to\infty}{\approx}\frac{{l_f\left( {l_f + 1} \right)}}{{2\mu_{3b}R^2 }} + E_{2b}^{\left({f}\right)},
\end{equation}
where $E_{2b}^{\left({f}\right)}$ is the two-body bound state (dimer) energies, and $l_f$ is the corresponding angular momentum of the third particle relative to the dimer. The subscript $f$ labels these recombination channels in ascending order, i.e., from high-to-low two-body bound state energies. In Figs. 1 and 2, red solid lines denote the three-body break-up channels (or entrance channels) whose asymptotic is described by
\begin{equation}\label{Uiasym}
\widetilde U_i  \left( R \right) \stackrel{R\to\infty}{\approx} \frac{{\lambda _i  \left( {\lambda _i  + 4} \right) + 15/4}}{{2\mu_{3b}R^2 }},
\end{equation}
where $\lambda_i \left( {\lambda_i  + 4} \right)$ is the eigenvalue of the grand angular momentum operator $\Lambda^2$ (here, $\lambda_i=0,4,6,8$..., where $\lambda_i=2$ is absent for symmetry considerations). The subscript $i$ labels three-body break-up channels in ascending order, i.e., from low-to-high eigenvalues $\lambda_i$.

As Ref. \cite{Esry2001, Dincao2005} points out, the asymptotic form of $\widetilde U_i$ determines the Wigner threshold laws for recombination, i.e., the low energy behavior of the recombination rate. A simple extension of the results of Ref.~\cite{Esry2001,Dincao2005} yields the Wigner threshold laws for all three-body channels as
\begin{equation}\label{WignerThreshold}
K_3^{\left( {f \leftarrow i} \right)}  \propto E^{\lambda _i }
\end{equation}
Our numerical results, however, show that Eq. (\ref{WignerThreshold}) only holds for the lowest entrance channel ($i=1$) while it fails to describe the threshold laws for higher incident channels $(i > 1)$. This is apparent from Figs. 3 and 5, which show our numerical calculations for recombination with positive and negative values of the scattering lengths, respectively.

In fact, Figs. 3 and 5 illustrate that for the partial recombination rate from the lowest three-body incidence channel ($\lambda_1=0$), the threshold behavior does follow the Wigner threshold law prediction, $K_3^{\left( {f \leftarrow 1} \right)}  \propto E^0$. However, for higher incident channels $\lambda_2=4$, $\lambda_3=6$, the threshold energy exponent is independent of $\lambda_{i}$ and recombination rates are only proportional to $E^1$. Therefore the low-energy suppression for higher three-body break-up channels is much weaker than what Wigner's threshold law would predict (see Eq. (\ref{WignerThreshold})). Note that we have used different line-styles to indicate the recombination rate for different incident channels, and use different color and thickness of lines for different final channels. The solid, dashed, dot, dash-dot and short dashed lines indicate recombination rate to different recombination channels: $f=1,2,3,4,5$. The thick black lines, thinner red lines, and thinnest blue lines indicate recombination rates from different incident channels: $i=1,2,3$.

Another important property that has emerged from our numerical calculations is that the branching ratio [Eq.~(\ref{ratio})] for the three-body recombination rates into different final channels are the {\it same} for the few lowest initial channel in the low collision energy limit (see Figs. 4 and 6). For instance, for the three different initial channels shown in Fig. 4, a case with positive scattering length, the branching ratio into the highest bound state is about $0.35$ for each of the three lowest incident channels throughout the energy range $E\lesssim 10\mu$K. Similar results are seen for the branching ratios at negative scattering lengths, as is documented by Fig. 6. Note, however, that the branching ratios for positive scattering lengths are not the same for $E>10\mu$K (see Fig. 4) while they remain the same for negative scattering length (see Fig. 6). This is a result of constructive interference effects that reduces the recombination probability for the most weakly bound recombination channel for positive scattering length, and it is related to the universal Efimov physics \cite{BraatenReview,Dincao2005}. In fact, such interference effect is only significant in the shallowest final channel. Hence, if the $f=1$ channel is excluded from the summation in the denominator on the right hand side of Eq.~(\ref{ratio}), the calculated branching ratio between the deep bound states should be the same for the whole energy range considered, as the inset of Fig. 4 shows.

Both the branching ratio properties uncovered in the present numerical exploration and the deviations from the recombination Wigner threshold laws can be understood using the analytical model developed in the next section. As we will see, these results are driven simply by the strong long-range coupling between the three-body incident channels; this analysis gives further insight into the pathways controlling three-body recombination.

\begin{figure}[t]
\includegraphics[width=0.5 \textwidth]{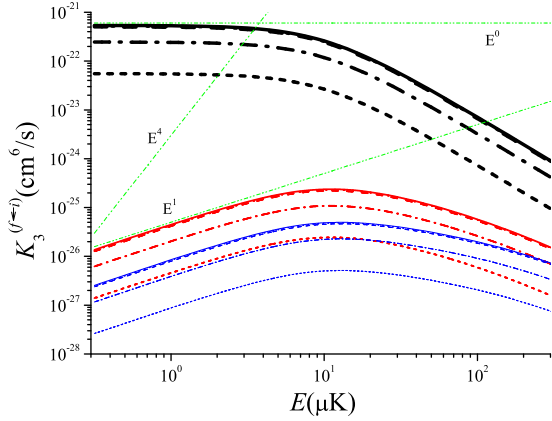}
\caption{(Color online) Same as Fig.3 but for the negative scattering length case.}
\end{figure}

\begin{figure}
\includegraphics[width=0.5 \textwidth]{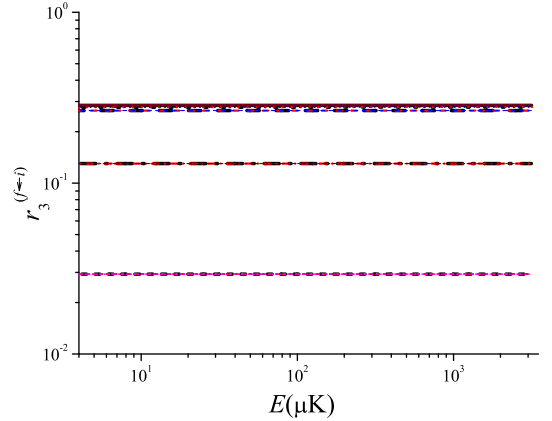}
\caption{(Color online) Same as Fig.4 but for the negative scattering length case.}
\end{figure}

\section{Dominant recombination pathways}
The extensive numerical three-body recombination rates presented in the preceding section are now interpreted in order to extract the important recombination pathways.  Once these are identified, the surprising low-energy threshold behavior and  the branching ratio regularities cited above become clear.

Our model consists of carrying out a perturbation expansion of the S-matrix and then associating each term to an specific pathway. As a first step, Eq. (\ref{RadialEq}) is recast in matrix form as

\begin{equation}
\left[ {\underline {T_R }  + \underline W - E\underline 1 } \right] \underline F = 0,
\end{equation}
where
\begin{equation}
\left( {T_R } \right)_{\mu \nu }  =  - \frac{1}{{2\mu_{3b}}}\frac{{d^2 }}{{dR^2 }}\delta _{\mu \nu }
\end{equation}
and
\begin{equation}\label{Wmat}
W_{\nu \mu }  = U_\nu  \delta _{\nu \mu }  - \frac{1}{{2\mu_{3b}}}\left[ {2P_{\nu \mu } \frac{d}{{dR}} + Q_{\nu \mu } } \right].
\end{equation}
The off-diagonal terms of $\underline W$ are treated perturbatively, suggesting that the hyperradial Green's function matrix should be defined as the solution of
\begin{equation}\label{GreenEq}
\left( {\underline T_R  + \underline {\widetilde U} - E\underline 1} \right)\underline G\left( {R,R'} \right) = \delta \left( {R - R'} \right)\underline 1
\end{equation}
where $\underline {\widetilde U}$ is the diagonal submatrix of $\underline W$, whose matrix elements coincide with $\widetilde U_\nu$ in Eqs. (\ref{Ufasym}) and (\ref{Uiasym}). One can, therefore, write the hyperradial Green's function as
\begin{equation}
\underline G \left( {R,R'} \right) = - \pi i\underline f \left( {R_ <  } \right)\underline h ^{\left(  +  \right)} \left( {R_ >  } \right),
\end{equation}
where $\underline f$ and $\underline h ^{\left(  +  \right)}$ are both diagonal matrices. The matrix elements of $\underline{f}$ are the solutions of Eq. (\ref{GreenEq}) regular at $R=0$, and the outgoing Hankel solutions $\underline h$ are given by
\begin{equation}
h^{\left(  +  \right)} \left( R \right) = f\left( R \right) + ig\left( R \right),
\end{equation}
where $g$ represent the corresponding irregular solutions. For the three-body break-up channels, since the centrifugal barriers are dominant, the regular and irregular energy-normalized basis pair $f_i$ and $g_i$ are well approximated in terms of Bessel functions as
\begin{subequations}
\begin{eqnarray}
f_i \left( R \right) &\approx& \sqrt {\mu_{3b}R} J_{\lambda _i  + 2} \left( {kR} \right), \\
g_i \left( R \right) &\approx& \sqrt {\mu_{3b}R} Y_{\lambda _i  + 2} \left( {kR} \right).
\end{eqnarray}
\end{subequations}
The above hyperradial Green's function can now be used to expand the S-matrix in a distorted-wave Born series,
\begin{equation}\label{Smatexpan}
\mathcal{S}_{fi}  = \mathcal{S}_{fi}^{\left( 0 \right)}  + \mathcal{S}_{fi}^{\left( 1 \right)}  + \mathcal{S}_{fi}^{\left( 2 \right)}  + ...,
\end{equation}
where
\begin{equation}
\mathcal{S}_{fi}^{\left( 0 \right)}  = \delta _{fi}  = 0,
\end{equation}
($f\ne i$). In Eq. (\ref{Smatexpan}), the first order expansion of the scattering matrix element is simply given by,
\begin{equation}\label{Smat1order}
\mathcal{S}_{fi}^{\left( 1 \right)}  = 2 \pi i \int_{0}^{\infty}{dR}{f_f \left( R \right)W_{fi}\left( R \right)} f_i \left( R \right),
\end{equation}
and the low energy behavior of the S-matrix elements can be easily determined by inspection. The integrand in Eq. (\ref{Smat1order}) is only significant at small values of $kR$ where $f_i \left( R \right) = \sqrt R J_{\lambda _i  + 2} \left( {kR} \right) \propto k^{\lambda_i  + 2}$. Therefore,
\begin{equation}\label{SFirstOrder}
\mathcal{S}_{fi}^{\left( 1 \right)}  \propto k^{\lambda_i  + 2}.
\end{equation}
In terms of the pathways, the first order S-matrix element is the probability amplitude to transit from the incident channel $i$ and then tunnel through the centrifugal barrier and scatter into recombination channels at short distances ($R\propto r_{0}$). Therefore, if the recombination process were solely described by this pathway, the low energy behavior of recombination would be given by
\begin{equation}\label{K3FirstOrder}
K_3^{\left( {f \leftarrow i} \right)}  = \frac{{192\pi^2\left( {2J + 1} \right)}}{{\mu_{3b}k^4 }}\left| {\mathcal{S}_{fi}^{\left( 1 \right)} } \right|^2  \propto k^{2\lambda _i }  = E^{\lambda _i },
\end{equation}
recovering the usual threshold laws from Wigner's analysis \cite{Esry2001,Dincao2005}.

The first-order result shown in Eq.~(\ref{K3FirstOrder}) for the low energy behavior of recombination fails, however, to explain our numerical coupled-channel results [see Figs. 3 and 5] implying that high order perturbation terms in Eq. (\ref{Smatexpan}) are crucial in order to determine the actual threshold laws. Hence we consider the second order partial-wave Born expansion, given by
\begin{equation}\label{SSecondOrder}
\mathcal{S}_{fi}^{\left( 2 \right)}  = - 2 \pi^2 \left( {I_1  + I_2 } \right),
\end{equation}
where,
\begin{subequations}
\begin{eqnarray}
I_1  &=& \sum\limits_{m\ne i\ne f} {\int_0^\infty  {dR} f_f \left( R \right)W_{fm} \left( R \right)f_m \left( R \right)} \nonumber \\ && \times \int_R^{\infty} {dR'} h_m^{\left(  +  \right)} \left( {R'} \right)W_{mi}\left( {R'} \right)f_i \left( {R'} \right), \\
I_2  &=& \sum\limits_{m\ne i\ne f} {\int_0^\infty  {dR}  f_f \left( R \right)W_{fm} \left( R \right)h_m^{\left(  +  \right)} \left( R \right)} \nonumber \\ && \times \int_0^R {dR'} f_m \left( {R'} \right)W_{mi}\left( {R'} \right)f_i \left( {R'} \right).
\end{eqnarray}
\end{subequations}
The first integral $I_1$ describes the quantum amplitude for a pathway in which the system coming inward in incident channel $i$ to first scatter into an intermediate state $m$ via a long-range coupling and then scatters to the final channel $f$ at short distances. $I_2$ describes the amplitude for a different second-order pathway for which the system first scatters into an intermediate state $m$ at short distances and then scatters into the final channel $f$ in a second step. Accordingly in our analysis, the most important pathway for all incident channels is the one associated with the $I_{1}$ term in Eq.~(\ref{SSecondOrder}), i.e., the pathways incorporated in $I_{2}$ are much more strongly suppressed in the low-energy limit.

Interestingly, the second-order correction for the S-matrix element associated with the lowest three-body incidence channel ($i=1$) can only produce a deeply-bound molecular channel or else an excited three-body continuum channel. In both cases, our analysis shows that these contributions are unimportant in the low-energy limit. Therefore, the threshold law for the lowest three-body channel is still given by Eq.~(\ref{K3FirstOrder}) [with $i=1$, $\lambda_{1}=0$]. For recombination events starting from excited three-body channels ($i>1$), however, the situation is different. In this case the dominant pathway is the one that involves the lowest three-body continuum channel as an intermediate channel ($m=1$), with a corresponding second order correction:
\begin{eqnarray}\label{Smat2order}
S_{fi}^{\left( 2 \right)}  && \approx -2 \pi^2 \int_0^\infty  {dRf_f \left( R \right)W_{f1} \left( R \right)f_1 \left( R \right)} \nonumber \\&& \times
\int_R^\infty  {dR'h_1^{\left(  +  \right)} \left( {R'} \right)W_{1i} \left( {R'} \right)f_i \left( {R'} \right)}.
\end{eqnarray}
The long-range coupling $W_{1i}$ between the lowest three-body break-up channel and a higher incident channel is dominated by the P-matrix element between the two channels. For different $i>1$, the P-matrix element $P_{1i}$ follows the same asymptotic behavior \cite{BurkeThesis}
\begin{equation}\label{Pmat}
P_{1i} \left( R \right) \propto \frac{1}{{R^2 }},~~~~~~~~~\left({R \to \infty}\right).
\end{equation}
Using the above equation and definition of $W_{\nu\mu}$ in Eq. (\ref{Wmat}), the integral in the second line of Eq. (\ref{Smat2order}) has the property that
\begin{equation}
\int_R^\infty  {dR'h_1^{\left(  +  \right)} \left( {R'} \right)W_{1i} \left( {R'} \right)f_i \left( {R'} \right)}  \propto k.
\end{equation}
The integral in the first line of Eq. (\ref{Smat2order}) is the same as Eq. (\ref{Smat1order}). Therefore, the second-order scattering-matrix element for the $i>1$ three-body break-up channels follows
\begin{equation}
\mathcal{S}_{if}^{\left( 2 \right)}  \propto k^{\lambda _1  + 2} k = k^3,
\end{equation}
which is larger than the first-order S-matrix for channels $i>1$ in the small $k$ limit [see Eq.~(\ref{SFirstOrder})]. Therefore, based on the discussions above, the threshold behavior of the partial recombination rate for any incident channel can be summarized as
\begin{equation}
K_3^{\left( {i \leftarrow f} \right)}  \propto
\left\{ {\begin{array}{*{20}c}
   {E^0,~~~~~~~~~i = 1},  \\
   {E^1,~~~~~~~~~i > 1},  \\
\end{array}} \right.
\end{equation}
which is consistent with our numerical results shown in Figs. 3 and 5.

The present analysis, therefore, demonstrates that the important recombination pathway for excited three-body incidence channels involves an intermediate transition to the lowest three-body incidence channel, controlled by a strong and long-range coupling between continuum channels [Eq.~(\ref{Pmat})]. This recombination pathway also explains why the relative recombination rate to reach the same final recombination channel from different incident three-body channels is the same. For ultracold collisions triggered from every excited three-body incidence channel ($i>1$), our analysis shows that the final state contribution for recombination is mainly controlled by the coupling between the lowest three-body break-up channel at short distances. Therefore, the corresponding relative final state contribution are independent of the initial excited three-body channel.
\section{Summary}
The methodology elaborated in this paper is capable of calculating recombination rate and, similarly, any other three-body scattering observable for systems that possess many two-body bound states. Our numerical study was performed for systems with up to 10 bound states, but it can be extended to larger problems with a good level of accuracy. Although our calculations for larger systems might be limited by memory usage and CPU time, our approach still allows for the analysis of increasingly more complex systems.

A key outcome is an understanding of the modified threshold laws for partial channel contributions to three-body recombination of three identical bosons with angular momentum $J=0$. Our analysis for the important recombination pathways reveals that the threshold behavior of the recombination rate for excited three-body incidence channels is significantly less suppressed at low energy than a simple generalization of the Wigner's threshold laws predicts. In addition, the branching ratio of recombination into any given final state $f$ is found to be the same for different incident channels.

\section{Acknowledgment}
This work was supported by NSF.  We thank Yujun Wang and Brett Esry for helpful discussions of numerous issues relevant to the present study.

\appendix
\section{R-matrix propagation method with traditional adiabatic method.}
The present model described in the main text uses the traditional adiabatic approach combined with an R-matrix propagation method for large values of $R$, where the $\underline P$ and $\underline Q$ matrices are smooth functions of $R$. One advantage of using this representation is that instead of calculating the values of the $\underline P$ and $\underline Q$ matrices at every mesh point in $R$, we can solve the hyperangular part of the Hamiltonian at relatively fewer grid points. (The number of grid points is generally set by the characteristic wave length associated with the collision energy.) This, therefore, allows the use of interpolation and/or extrapolation methods in order to generate the required much denser grid without memory storage problems. This Appendix describes the approach in more detail.

In the traditional method, $\psi_{\nu '}$ is expanded as
\begin{equation}\label{PQExpand}
\psi _{\nu '} \left( R \right) = \sum\limits_{j\nu } {c_{j\nu,\nu ' } \pi _j \left( R \right)\Phi _\nu  \left( {R;\Omega } \right)}.
\end{equation}
A comparison of Eq. (\ref{PQExpand}) with Eq. (\ref{SVDExpand}) shows that the main differences between the traditional adiabatic method and the SVD method derive from using different three-body numerical basis sets. (Notice that in Eq. (\ref{PQExpand}), the $\Phi _\nu  \left( {R;\Omega } \right)$ are channel functions evaluated at $R$, while in Eq. (\ref{SVDExpand}), the $\Phi _\nu  \left( {R_j;\Omega } \right)$ are channel functions evaluated at $R_j$.) However, one can show that the expansion coefficients $c_{j\nu, \nu' }$ are the same for the two expansions if $\Phi _\nu  \left( {R;\Omega } \right)$ is smooth so that the DVR approximation can be applied,
\begin{equation}\label{DVRIntegral}
\int {dR} \pi _i \left( R \right)\Phi _\nu  \left( {R;\Omega } \right)\pi _j \left( R \right) \approx \Phi _\nu  \left( {R_i ;\Omega } \right)\delta _{ij}.
\end{equation}
Eq. (\ref{DVRIntegral}) implies that the traditional adiabatic method and the SVD method are equivalent within the DVR approximation. Therefore, when the $\underline P$ and $\underline Q$ matrices change rapidly and are hard to evaluate numerically, it is highly advantageous to choose the SVD method; when the $\underline P$ and $\underline Q$ matrices are smooth, however, the traditional method is simpler and benefits from lower memory storage requirements.

Next, the details of the traditional approach are elaborated and  the R-matrix propagation from a point $b_1$ to another point $b_2$ is explained. Insertion of Eq. (\ref{PQExpand}) into Eq. (\ref{Fmatrix}) yields
\begin{eqnarray}\label{Fmatrixelement}
F_{\nu \nu '} \left( {b_1 } \right) &=& \sum\limits_{j } {c_{j\nu, \nu'} \pi _j \left( {b_1 } \right)},\\
F_{\nu \nu '} \left( {b_2 } \right) &=& \sum\limits_{j } {c_{j\nu, \nu'} \pi _j \left( {b_2 } \right)},\nonumber\\
\widetilde F_{\nu \nu '} \left( {b_1 } \right) &=& \sum\limits_\mu  {\left[ {\delta _{\nu \mu } F_{\mu \nu '} '\left( {b_1 } \right) + P_{\nu \mu } \left( {b_1 } \right)F_{\mu \nu '} \left( {b_1 } \right)} \right]},\nonumber\\
\widetilde F_{\nu \nu '} \left( {b_2 } \right) &=& \sum\limits_\mu  {\left[ {\delta _{\nu \mu } F_{\mu \nu '} '\left( {b_2 } \right) + P_{\nu \mu } \left( {b_2 } \right)F_{\mu \nu '} \left( {b_2 } \right)} \right]}.\nonumber
\end{eqnarray}
As the next step, rewrite Eq. (\ref{SchrodingEq}) in the basis of Eq. (\ref{PQExpand}) in the matrix form of Eq. (\ref{matrixform}), with the matrix elements
\begin{widetext}
\begin{eqnarray}
H_{i\mu ,j\nu }  &=& \frac{1}{{2\mu_{3b}}}\int_{b_1 }^{b_2 } {\pi _i '\left( R \right)\pi _j '\left( R \right)dR} \delta _{\mu \nu }  + \left[ {U_\nu  \left( {R_j } \right)\delta _{\mu \nu }  + \frac{{P^2_{\mu \nu } \left( {R_j } \right)}}{{2\mu_{3b}}}} \right]\delta _{ij} \nonumber \\
 && - \frac{1}{{2\mu_{3b}}}\int_{b_1 }^{b_2 } {\pi _i \left( R \right)P_{\mu \nu } \left( R \right)\pi _j '\left( R \right) - \pi _i '\left( R \right)P_{\nu \mu } \left( R \right)\pi _j \left( R \right)dR},\\
L_{i\mu ,j\nu }  &=& \left. {\frac{1}{{2\mu_{3b}}}\left[ {\pi _i \left( R \right)\delta _{\mu \nu } \pi _j '\left( R \right) + \pi _i \left( R \right)P_{\mu \nu } \left( R \right)\pi _j \left( R \right)} \right]} \right|_{b_1 }^{b_2 }.
\end{eqnarray}
\end{widetext}
Use of these matrix elements and replacing $a1$($a2$) with $b1$($b2$), the same procedure as Eqs.~(\ref{diagH}-\ref{RmatProb}) accomplishes the matrix propagation.


\begin{thebibliography}{99}
\bibitem{addnote} For the $f$'th recombination channel, $l_f$ is given by Eq.~(\ref{Ufasym}), and $k_f=\sqrt{2\mu_{3b}\left({E-E_{2b}^{\left({f}\right)}}\right)}$. For the $i$'th three-body break-up channel, $l_i=\lambda_i+3/2$, and $k_i=\sqrt{2\mu_{3b}E}$.
\end{thebibliography}
\end{document}